\documentstyle[aps,multicol,prl]{revtex}
\renewcommand{\narrowtext}{\begin{multicols}{2}
\global\columnwidth20.5pc}
\renewcommand{\widetext}{\end{multicols}
\global\columnwidth42.5pc} 
\begin{document}

\preprint{Submitted to PRL}

\draft
\title{Level statistics in a two-dimensional system with strong
spin-orbit coupling}

\author{ A.~P.~Dmitriev and V.~Yu.~Kachorovskii}

\address { A.~F.~Ioffe Physical-Technical Institute, 26 Polytechnicheskaya str.,
St.\ Petersburg 194021, Russia\\ e-mail:
dmitriev.vip1@pop.ioffe.rssi.ru} 

\maketitle
\begin{abstract}

{We study level correlations in a two-dimensional system with a
long-range random  potential and strong spin-orbit (SO)  
splitting $\Delta_{so}$ of the
spectrum.  
The   level correlations 
for  sufficiently large $\Delta_{so}$ are shown to be described by 
  orthogonal Wigner-Dyson statistics, in contrast 
  to the common point of view that  a system with the strong  SO  coupling belongs to 
the  symplectic statistical ensemble.
We demonstrate also that in 
a wide energy interval the level statistics is
completely determined by  transitions 
between  two branches of the split spectrum.  
A sharp resonance  is obtained in the two-level correlation function 
 at  energy equal to $ \Delta_{so} . $ }

\end{abstract}
\narrowtext
\vspace{0.5cm}

The advancement of fabrication of nanostructure  devices in the last two decades has 
initiated  great interest to properties of small mesoscopic
systems (see Ref.~\cite{imry} for review).
One of the most frequently studied characteristics of the mesoscopic samples 
is the two-level correlation function $ R(\epsilon) = \langle
\rho(E+\epsilon)\rho(E)\rangle/\rho^2 -1  $ which describes
fluctuations of the density of states $\rho(E ).$
Here $\rho = \langle \rho(E)\rangle$ and $\langle ...\rangle$ 
denotes averaging over  realizations of the random 
potential.
The problem of level correlations in small  disordered conductors
was  
first discussed in \cite{gorel}, where it was suggested that the
distribution of 
 energy levels is determined by one of the three types of the universal
Wigner-Dyson statistics \cite{wig}.
Later on this idea was confirmed by rigorous calculations  \cite{ef} of 
$R(\epsilon)$. It has been proved  that the universal
Wigner-Dyson distribution is 
realized at 
small energies $\epsilon \ll E_c$, where $E_c \sim \hbar/\tau_c$  
is the Thouless energy 
 (here $\tau_c \sim L^2/D $, $L$ is the size of the system and  $D$ is the diffusion 
coefficient).  
At larger
energies ($\epsilon \gg E_c $) the level statistics is nonuniversal
and depends  on the dimensionality of the system \cite{shkl}.
The most interesting is the $2d$ case in which  the  calculation 
of $R(\epsilon)$ in the standard one-loop approximation
gives  a zero result \cite{shkl}.
  In this situation the level
statistics  is governed by weak localization effects \cite{krav}
and is sensitive to the
boundary conditions and topology of the surface \cite{yud}.

In this paper we study level correlations in 
 a $2d$ system with strong SO 
spectrum splitting $\Delta_{so}$, which originates from an asymmetry of the confining 
potential of the quantum well\cite{rashba}. The enhanced interest to the systems 
 with strong SO coupling   was initiated   by recent remarkable experiments of 
Kravchenko {\it et al.} \cite{kravch}  which clearly demonstrated the existence 
of a metal-insulator transition in $2d$ systems. The importance of the 
SO spectrum splitting
 for the 
transition was first conjectured in Ref.~\cite{pud}.
This idea was strongly supported by very recent experimental data \cite{papa,yaish}. 
 
Here, we 
show that strong SO spectrum splitting leads to unexpected effects 
 in the statistics of levels
in a $2d$ disordered system. 
A common belief is that the type of the level statistics in  
a system with the  SO coupling depends only on the relation between  
the spin relaxation time $\tau_{so}$ 
and $\tau_c .$ 
For the case of very small 
 splitting, when 
$\tau_{so} > 
 \tau_c ,$  spin degrees of freedom do not play any role and the system belongs
 to the   orthogonal ensemble - one 
 of the three  universal Wigner-Dyson ensembles. 
 With increasing SO coupling, 
 $\tau_{so} $ becomes smaller than $\tau_c$,
 the spin relaxation comes
 into play and the system exhibits crossover to  symplectic ensemble. 
   This scenario  takes place  for the case of a short-range potential.
 As we show below, the system with long range disorder demonstrates {\it reentrance to
the  orthogonal ensemble} at sufficiently large values of $\Delta_{so}$. This surprising
result is  related to   
strong suppression of the transitions between two
branches of the SO split spectrum in the case 
$\Delta_{so} \gg \hbar v_F/d $ \cite{my} (here
$d  $ is the potential correlation length, $v_F$ is the Fermi velocity). 
As a consequence, the system is divided into two independent subsystems, the spin degrees of
freedom are effectively frozen out, so that  in the limit of very large $\Delta_{so}$ we have   
two independent  orthogonal ensembles.
We will demonstrate also  that long-range nature of the impurity potential is of 
crucial importance  for the level statistics in the nonuniversal region of energies $\epsilon \gg E_c$. 
We argue that in a wide energy interval in the nonuniversal region
the level correlations are 
 completely governed by the rate of transition  between  two branches of the 
spectrum 
and   
that a narrow resonant peak arises in 
$R(\epsilon)$
 for $\epsilon = \Delta_{so} .$ 
 
    The term in the Hamiltonian which is
responsible for the SO splitting of the energy spectrum
\cite{rashba} is given by $\alpha({\bf 
\hat
\sigma}\times{\bf
k})\cdot {\bf n}$, where 
 ${\bf 
 \hat
 \sigma}$ are the Pauli matrices,
$\hbar {\bf k}$ is the electron momentum  and  ${\bf n}$ 
is the unit vector normal to the $2d$ layer.
A constant $\alpha$ characterizing the strength of the SO coupling 
  is proportional to the electric field arising due 
  to asymmetry of the confining potential of the quantum well. 
The value of $\alpha$ can be tuned 
 by applying an external electric field
perpendicular to the $2d$ plane \cite{papa}.  
  The energy spectrum
and spin wave-functions of an electron with the effective mass $m$ read:  
$E^{\pm}({\bf k})=\hbar^2 k^2/2m \pm \alpha k, \;
 \chi^{\pm}_{\bf k}= \left(1, \pm i e^{i \varphi_{\bf k}} 
\right)/\sqrt{2}.$
The spinors $\chi^{\pm}_{\bf k}$ depend on the polar angle $\varphi_{\bf
k}$ of the wave vector ${\bf k}$ and describe two states with 
spins polarized along the vectors $\pm ({\bf k}\times {\bf n})$,
respectively.  Thus the system is divided into two 
branches, ($+$) and ($-$) (see Fig.~1a) and the electron spin
 in each of the branches is rigidly connected to the momentum.

The long-range random impurity potential $U({\bf r})$ 
is assumed to 
be weak enough, so
that the  inequality $\tau \gg d/v_F$ holds, where 
$\tau \sim \tau_{tr}/(k_Fd)^2 $ is the elastic scattering time, ${k_F}$ 
is the Fermi 
wave-vector. 
  The presence of this potential leads to transitions
both within each branch and between different branches.  
The respective (intrabranch and interbranch) times of these transitions
are given by the following expression
 \begin{equation}
\frac{1}{\tau_{\mu \nu}}=\frac{2\pi}{\hbar} \int \frac{d^2 {\bf
k^\prime}}{(2\pi)^2} K_{\mu \nu}({\bf k}-{\bf k^\prime}) \delta[E^{\mu}({\bf k}) 
-
E^{\nu}({\bf k^\prime})], 
\label{tau*} 
\end{equation}
 where $K_{\mu\nu}({\bf k} -{\bf k^\prime})= | \langle \chi^{\mu}_{\bf k} | 
\chi^{\nu}_{{\bf k^\prime}
 } \rangle |^2 K(|{\bf k}-{\bf k^\prime} |) $, indices $\mu$ and $\nu$ denote 
the type of
the branch ($+$ or $-$), and the function 
$K(k)$ is the
Fourier-transform of the potential correlation function $K(r) =\langle U({\bf r})U(0) \rangle$. 
The function $K(r)$ falls off on the 
scale of $d$. 
For the case when the spin-orbit splitting
$\Delta_{so} = 2 \alpha k_F$ is not too large 
($\Delta_{so} \ll E_F$, where $E_F$ is the Fermi energy),
 we can set $\tau_{++}=\tau_{--}=\tau$,
$\tau_{+-}=\tau_{-+}=\tau_{*}$.  The minimal transferred momentum
needed for the transition between the branches 
is equal to $2 m \alpha/\hbar .$
This momentum should be compared with $\hbar/d$.
If the inequality
 $\Delta_{so} \gg \hbar v_F/d $  
   is fulfilled, then the interbranch transitions are
suppressed compared to transitions within one branch (note that 
this inequality 
implies also that $ \Delta_{so} \tau \gg \hbar $).  In particular,
for the case of the potential created by ionized impurities located at
distance $d$ from the $2d$ layer, the correlation function $K(q)\sim
\exp(-2qd)$ and the interbranch transition time is given by
\begin{equation} 
\tau_*=4\tau  (k_F d)^2 \exp\left(\frac{2 \Delta_{so}
d}{\hbar v_F}\right)\: \gg \tau_{tr} .  
\label{tauexp} 
\end{equation}
 The factor
$(k_F d)^{2}$ in this expression is due to the orthogonality of
the spinors corresponding to different branches with the same direction
of momenta.  
Note that $\tau_* \sim \tau_{tr}$ for the case of
 a short-range potential ($K(q)= const $).

Expressing the density of states  in the usual way in terms of exact Green 
functions
$
\rho(E) = (i/2\pi L^2) {\rm Tr}( \hat G^R(E) - \hat G^A(E)),
$ 
 we can write
 the two-level 
correlation function  as 
$R(\epsilon)= (\Delta^2 /8 \pi^2) {\rm Re} 
  \langle {\rm Tr} \hat G^R (E +\epsilon)
{\rm Tr} \hat G^A ( E )\rangle,$
where $\Delta = 2\pi \hbar^2/ m L^2 $ is the mean level spacing  
and the trace is calculated over spatial and spin variables.
 The main contribution to the  
function $R(\epsilon)$ comes from one-loop 
diagrams containing  two diffuson or cooperon propagators, 
shown in  Fig.~2 in  the so-called Hikami representation \cite{hik}.
 In what follows 
we consider the case of 
zero magnetic field, so that the cooperon contribution is
equal to the diffuson one. 
This allows us to focus on  the calculation of the diffuson contribution only.  
  To separate the   two branches of the 
spectrum,  we use the 
following representation for the averaged Green 
functions 
$\langle\hat G^{R,A}({\bf k}, E) \rangle =
\sum_{\mu=\pm} G^{R,A}_\mu ({\bf k}, E)\hat P_\mu({\bf k}),$
where $ \hat P_\mu({\bf k }) = |\chi^{\mu}_{\bf k} \rangle \langle 
\chi^{\mu}_{\bf k}|$ is the projection  operator  to
the branch $\mu$.
The functions $G^{R,A}_\mu$ are represented by edges of the "Hikami boxes" in
Fig.~2.  
It is convenient to distinguish 
two channels:  the
 symmetric channel with 
 $ \mu=\nu , \:\: \mu^\prime = \nu^\prime $ 
   and antisymmetric channel with $ \mu = - \nu , \:\: \mu^\prime = - 
 \nu^\prime $ 
 (see Fig.~2).
 We will first concentrate on the region of small energies
($\epsilon \tau_{tr} \ll \hbar$), where the symmetric contribution
dominates.
The  diffuson propagator in symmetric channel depends on 
 two indices $\mu$ and $ \mu^\prime$
only: $D_{\mu \mu^\prime}^{\mu \mu^\prime}=D_s^{\mu \mu^\prime}$.
This propagator   corresponds
to the contribution of two coherent wave ("particle" and "hole" 
waves) which belong to the same branch (see Fig.~1b) in  
any part of the diffusive trajectory.
 The value $D_s^{\mu \mu^\prime}$ is determined from the
 system of four equations 
($\mu=\pm, \mu^\prime = \pm$) depicted graphically in Fig.~3,
where $ K_s^{\mu \mu^\prime}({\bf k},{\bf k^\prime},{\bf q}) =
K(|{\bf k}-{\bf k^\prime}|) 
\langle \chi^\mu_{\bf k} |\chi^{\mu^\prime}_{\bf k^\prime} \rangle
\langle \chi^{\mu ^\prime}_{{\bf k^\prime} +{\bf q}} |\chi^{\mu}_{{\bf k}+{\bf 
q} } 
\rangle. $
These equations  should be solved together with 
the system of equations for $G^{R,A}_\mu ({\bf k},E)$ 
written in the framework of  Self-Consistent Born Approximation.
In the diffusion approximation, 
the solution 
is given by 
$D^{++}({\bf q},\epsilon) = D^{--}({\bf q},\epsilon) =S_+\hbar/\pi\rho\tau^2$ 
and 
$D^{+-}({\bf q},\epsilon) = D^{-+}({\bf q},\epsilon)=S_- \hbar/\pi\rho\tau^2 $
\cite{kk}, 
where functions
\begin{equation} 
S_{\pm}({\bf q},\epsilon) =\frac{\hbar}{2} \left(   
\frac{1}{\hbar Dq^2 - i \epsilon} \pm \frac{1}{\hbar Dq^2 - i \epsilon 
+\Gamma^*}
\right)
\label{ds1}
\end{equation} 
obey the
set of  equations \cite{my}     
\begin{eqnarray}
\frac{\partial S_+ }{\partial t} - D \Delta S_+ &=& \delta({\bf
r})\delta(t) + \frac {S_- - S_+}{\tau_*},
\nonumber \\ 
\frac{\partial S_- }{\partial t} - D \Delta S_-
&=& \frac {S_+ - S_-}{\tau_*}.
\label{D} 
\end{eqnarray}
Here  $D =v_F \tau_{tr}/2 $ is the diffusion coefficient, $\Gamma_* = 
2\hbar/\tau^* $. 
Note that in the case of a short-range potential $\tau^* \sim \tau_{tr} $ and the  
term 
$(\hbar Dq^2 - i \epsilon + \Gamma_*)^{-1}$ can be neglected in both $S_+$ and
  $S_-$.
The contribution of the symmetric
channel to the two-level correlation function  is 
given by 
\begin{eqnarray}
R_s(\epsilon)=\frac{\Delta^2}{2\pi^2\hbar^2}\:\:Re\: \sum_{\bf 
q}\left(S_+^2({\bf q},\epsilon) +
 S_-^2 ({\bf q},\epsilon) \right ) = 
\nonumber \\
\frac{\Delta^2}{4\pi^2} Re \sum_{\bf q}\left( \frac{1}{( \hbar Dq^2 - 
i\epsilon)^2 } 
+ 
\frac{1}{( \hbar Dq^2 - i\epsilon + \Gamma_*)^2 }    \right).
\label{korr1}
\end{eqnarray}
The first term in the r.h.s of this expression is the
well-known Altshuler-Shklovskii 
result \cite{shkl}. The second term is related to the
interbranch transitions.

First we discuss the universal region of energies,
$\epsilon \ll E_c $. 
For $\Gamma_* \ll E_c $ \cite{gamma} 
one should 
keep only the term $ {\bf q} = 0 $
  in  Eq.~(\ref{korr1}), which yields
\begin{equation}
R_s(\epsilon) = 
 \frac{\Delta^2}{4\pi^2} \left(- \frac{1}{\epsilon^2} + \frac{\Gamma_*^2 
-\epsilon^2}
 {(\epsilon^2 + \Gamma_*^{2})^2}  \right)  .
\label{uni}
\end{equation}
A crossover between two  ensembles is clearly seen.
  For $\epsilon \ll \Gamma_* $ the first term in Eq.~(\ref{uni}) dominates. 
In this case
during the  time $\hbar/ \epsilon $
many interbranch transitions 
take place, the  system is ergodic and we obtain
the standard
result \cite{shkl} for the envelope of the symplectic Wigner-Dyson distribution. 
For $ \Gamma_* \ll \epsilon \ll E_c $ 
the time $ \sim \hbar/ \epsilon $ is not enough for interbranch transition
  and we have two independent 
orthogonal
ensembles. Thus, we see that the system demonstrates reentrance to the
orthogonal ensemble at large $\Delta_{so},$ when $\Gamma_*$ becomes sufficiently    
small.

In the nonuniversal region $\epsilon \gg E_c$, where the summation in 
Eq.~(\ref{korr1}) can be replaced 
by the integration, the contribution of the first term
in Eq.~(\ref{korr1}) is equal to zero
(for periodic boundary conditions) and the 
level correlations are governed by the interbranch transitions only: 
\begin{equation}
R_s(\epsilon)= \frac{\Delta}{4\pi g} \:\: \frac{ \Gamma_*}{\epsilon^2 + 
\Gamma_*^2 }.
\label{rs}
\end{equation}
Here $g = \pi k_F l_{tr} \:\: $ is the conductance in  units of  $e^2/\pi 
h$.

Compare now Eq.~(\ref{rs}) with the 
weak localization correction to $R(\epsilon)$  for the symplectic ensemble 
\cite{krav}
$R_{wl}(\epsilon)= \Delta/8g^2\epsilon $. The inequality 
$R_s(\epsilon) \gg R_{wl}(\epsilon)$ is fulfilled in the energy interval
${\rm max}(E_c, \Gamma_*/g) \ll \epsilon \ll g\Gamma_* $. 
It can be shown \cite{top} that the boundary-induced and 
topological contributions \cite{yud} are also small
compared with Eq.~(\ref{rs}). 
Thus, we 
see
that in the nonuniversal region the
interbranch transitions completely  govern level correlation in a wide energy 
interval. 

Now we will show that in the region of large energies $|\epsilon| \approx 
\Delta_{so} $ 
a  narrow resonant peak appears in the two-level correlation function.
The physical
 origin of this peak is this:  the exact energy levels lying
 near the energy $E$ in the lower branch ($-$) are correlated with
 the exact energy levels lying near the energy $E + \Delta_{so} $
 in the upper branch ($+$) (see Fig.~1c). The resonance contribution
 to $R(\epsilon)$ comes from the diagram in Fig.~2  
with $\mu = - \nu $ and $\mu^\prime = - \nu^\prime$ (antisymmetric channel).
The diffusons in antisymmetric channel 
depends on   indices ${\mu}$ and
${\mu^\prime}$ : 
$D^{\:\mu\: \mu^\prime}_{-\mu -\mu^\prime} = D_a^{\mu\mu^\prime}$. 
  The equations for $D_a^{\mu \mu^\prime}$ are 
 obtained from equations for $D_s^{\mu \mu^\prime}$ by 
changing  the sign of 
all
 the indices related to the line corresponding $G^A$ (lower line in 
Fig.~3).
 The dashed line represents now the
 function 
  $ K_a^{\mu \mu^\prime}({\bf k},{\bf k^\prime},{\bf q}) =
K(|{\bf k}-{\bf k^\prime}|) 
\langle \chi^\mu_{\bf k} |\chi^{\mu^\prime}_{\bf k^\prime} \rangle
\langle \chi^{-\mu ^\prime}_{{\bf k^\prime} +{\bf q}} |\chi^{-\mu}_{{\bf k}+{\bf 
q} } 
\rangle. $ In the case $\epsilon \approx \Delta_{so}$ the main contribution is represented
by 
$ D^{++}_a({\bf q},\epsilon) = ( \hbar^2 / \pi \rho \tau^2)
/( \hbar D q^2  + {\tilde \Gamma } -i 
  [\epsilon - \Delta_{so}]  
),$
where ${\tilde \Gamma} = \hbar/{\tilde \tau} $ and
\begin{equation}
\frac{1}{{\tilde \tau}} =
 \frac{m }{4 \pi \hbar^3 \Delta_{so}^2} \int \frac{p dp }{(2\pi)^2 }
 K_{tr}^2(p) \sim \frac{v_F}{d} \left( \frac{\hbar }{ \Delta_{so} \tau_{tr} 
}\right)^2.
\label{tau1}
\end{equation}
Here
$
K_{tr}(k- k^\prime)=\int  K(|{\bf k}-{\bf k^\prime}|)(1- \cos(\varphi))d\varphi
$ and
$\varphi$ is the angle between ${\bf k}$ and $\bf k^\prime$.
At large spectrum splitting the   inequalities $\tau_{tr} \ll {\tilde{\tau}} \ll 
\tau^*$ hold.
Now we 
present
 qualitative arguments 
clarifying the physical origin   of $\tilde{\tau}$ which
plays the role 
of the phase breaking time in the antisymmetric channel. 
The appearance of a resonance peak in $R(\epsilon)$ is a consequence of coherence
of the waves   ($+$) and 
 ($-$),  which for  
 $\epsilon=\Delta_{so}$ have the same momenta (see Fig.~1c). 
One might think that the 
coherence is destroyed by interbranch transitions only.
However, in the second 
order of perturbation theory 
there exists a virtual process which breaks the coherence.
In this process, 
the wave which in the initial and final state belongs to the branch ($+$), 
in the intermediate virtual  state belongs to the branch ($-$). This process 
leads to a small correction  to the effective random potential 
 which can be estimated as $W \sim U^2/\Delta_{so} (k_F d)^2 $ (here $U$ is the 
amplitude of random potential and
the factor $(k_F d)^{- 2} $ is of the same origin as in  Eq.~(\ref{tauexp})). 
 The other wave, 
in  the initial and final state belonging to the branch ($-$), 
in the intermediate virtual  state belongs to the branch ($+$). It is evident 
that the correction to the potential  $W^\prime$ has the same absolute value but the different 
sign 
$W^\prime = - W$. The sign difference is related to the fact that the energy of 
the virtual state
lies  in first process above, and in the second process below   energy of the initial 
state.
As a result, two waves propagate in slightly different effective potentials $U + W$ and $ 
U - W$.
On passing the potential correlation length $d$ they acquire the phase difference 
$\delta \phi \sim W d /v_F \hbar $. The phase breaking time is then easily 
estimated as 
$\tilde{\tau}^{-1} \sim (\delta \phi)^2 v_F/d $. Combining these formulas one 
can obtain the estimate 
for $\tilde{\tau} $ given in the r.h.s. of Eq.~(\ref{tau1}). 
 The resonance contribution to $R(\epsilon)$  reads     
 \begin{eqnarray}
  &R_a&(\epsilon) = 
\frac{\Delta}{8 \pi g} \frac{{\tilde \Gamma}}{( \epsilon -\Delta_{so})^2 
+ {\tilde \Gamma}^2 }~, \:\:\:\:  {\tilde \Gamma} 
\gg E_c~; 
\\
\nonumber
 &R_a&(\epsilon) = \frac{\Delta^2}{8\pi^2}  
\frac{{\tilde \Gamma} ^2 - ( \epsilon - \Delta_{so} )^2}
 {[ (\epsilon - \Delta_{so})^2 + {\tilde\Gamma}^2 ] ^2}~, 
 \:\: \:\: \Delta \ll \tilde{\Gamma} \ll E_c~.  
\label{rt}
\end{eqnarray}

 We conclude that a $2d$ system with a long-range random potential 
 and  strong SO splitting of the spectrum
  exhibits a 
  {\it reentrance to the orthogonal ensemble} for sufficiently large values of
  $\Delta_{so}.$
 A weak  interaction between
  two sets of random levels corresponding to  two branches of the split spectrum
   leads to the appearance of two effects: 
   (i)~in a wide energy interval the level statistics is completely governed by interbranch transitions;
    (ii)~a sharp resonance arises in the two-level correlation function for energies corresponding   
to the spectrum splitting.

The authors are grateful to M.~I.~Dyakonov,  I.~V.~Gornyi, A.~D.~Mirlin, and D.~G.~Polyakov
 for useful discussions. 
This work was supported by 
RFBR (99-02-17093, 99-02-17110, 96-15-96392), Grant for Young
Researchers of the RAS,  INTAS (grants
96-196, 97-1342), and by Program  "Physics of Solid State
Nanostructures" 
(grant 1001).
\\
\\
\\
\\

{\bf FIGURE CAPTIONS}

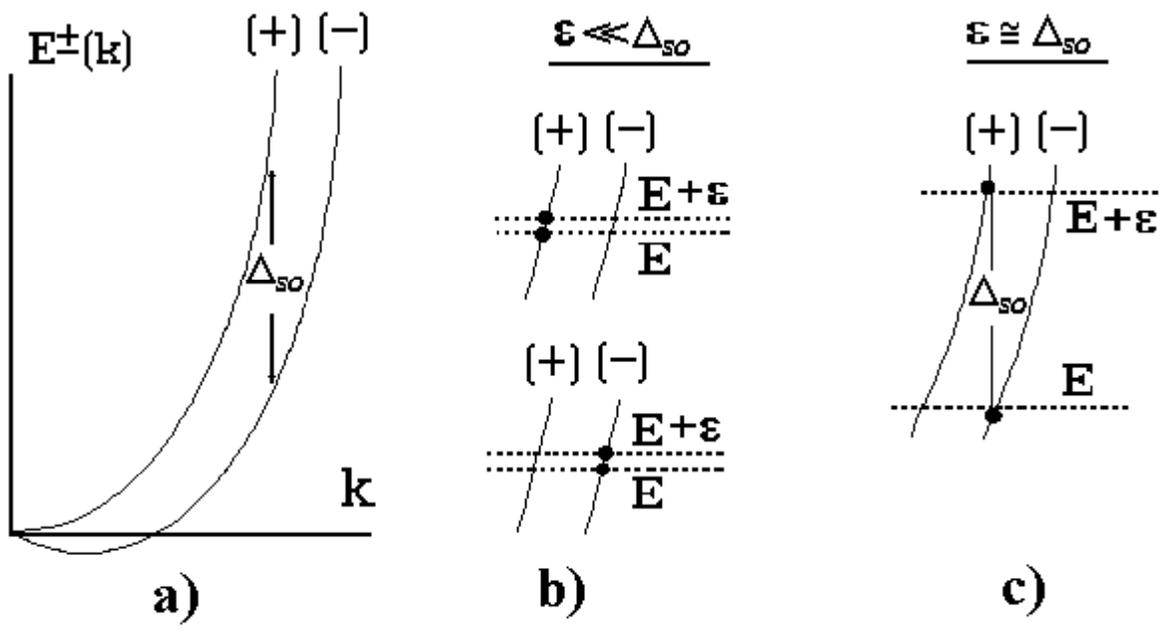
{\bf Fig.~1} 
(a)~Spin-orbit split energy spectrum;
(b)~symmetric contribution to the diffuson (two coherent waves 
are depicted by black dots); (c)~antisymmetric contributions to the 
diffuson. 

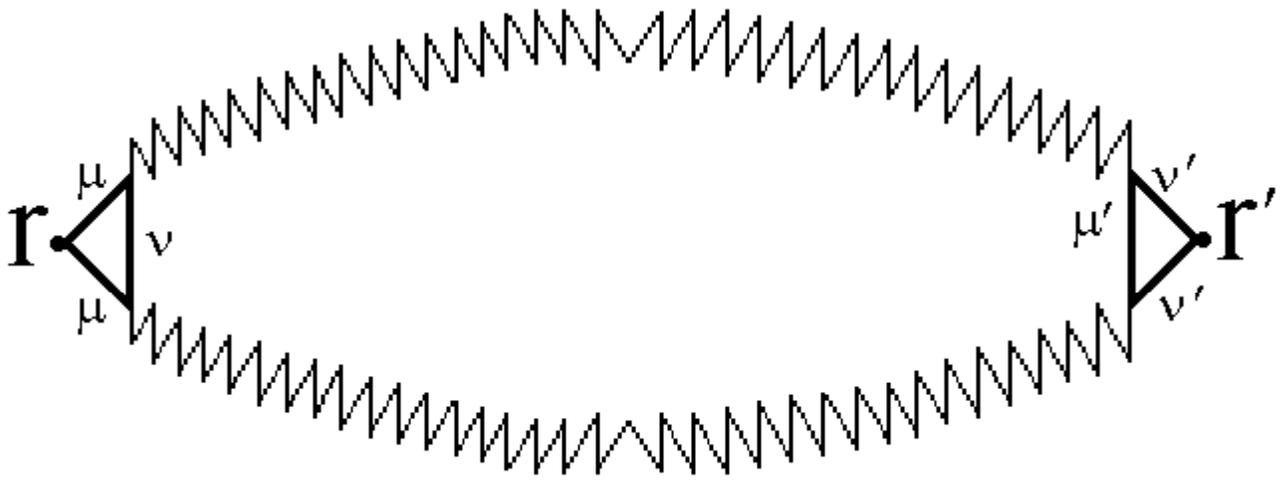
{\bf Fig.~2} One-loop diagram contributing to $R(\epsilon).$  

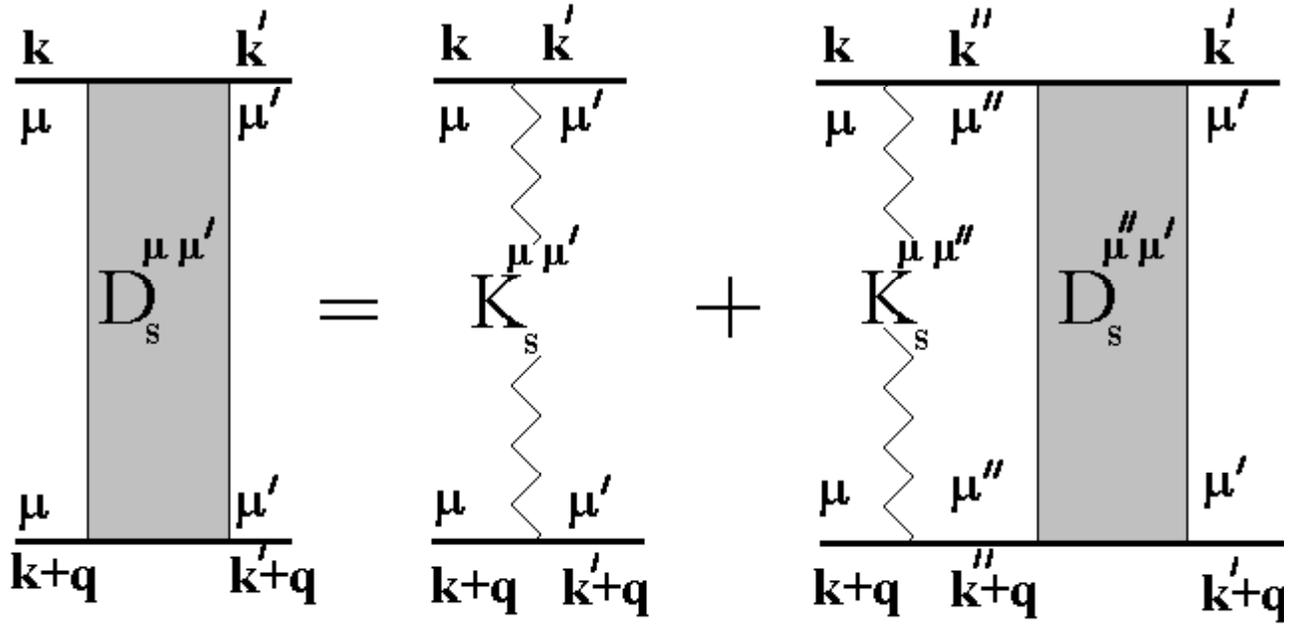
{\bf Fig.~3} Diffuson equation in the symmetric channel.

\widetext

\begin{references}
\bibitem{kk} Note that the diffuson in our case depends on  $|{\bf k}|$ 
and $|{\bf k^\prime}|$ (See Fig.\ 3). Eq.\ (\ref{ds1}) is valid
for $|{\bf k}|=k_\mu, |{\bf k^\prime}| 
=k_{\mu^\prime} $, where the values of $k_\mu$ are
determined from the conditions $E =E^+(k^+)=E^-(k^-)$.
\bibitem{gamma}  For $\Gamma_* \gg E_c$  
the interbranch contribution  to 
$R_s(\epsilon)$ in universal region  is  small 
  and leads 
 only to small correction to the   symplectic Wigner-Dyson distribution.
\bibitem{top} The topological tail 
 in $R(\epsilon)$  (for the case $p=0$ in notation of \cite{yud}) 
$R_{top} = - \Delta^2/ (12\pi^2\epsilon^2)$ is smaller than $R_s(\epsilon)$ if
$\epsilon \gg (\Gamma_* E_c)^{1/2}$. 
The boundary-induced  contribution  
$R_b(\epsilon) = -  \sqrt{2} \Delta^{3/2}  L_b /64 \pi g^{1/2} \epsilon^{3/2} 
L $ (here $ L_b $ 
is the length of the boundary and $L$ is the system size)  is negligible 
for $ (\Gamma_*^2 E_c L_b^2/L^2)^{1/3} \ll \epsilon \ll
\Gamma_*^2 L^2/(E_c L^2)$, 
provided that $ \Gamma_* \gg E_c L_b^2/L^2 .$
\\ 
\\ 
\\ 
\\ 
\\ 
\\ 
\\ 
\\ 
\\ 
\\ 
\\ 
\\ 
\\ 
\\ 
\\ 
\\ 
\\ 
\\ 
\\ 
\\ 
\\ 
\\ 
\\ 
\\
\\
\\
\\
\\
\end{references}
\end{document}